\pdfoutput=1 
\documentclass[
superscriptaddress,
 aps,
reprint
]{revtex4-1}

\usepackage{graphicx}
\usepackage{dcolumn}
\usepackage{bm}
\usepackage{amsmath}
\usepackage{color}
\usepackage{float}
\usepackage{array,url,kantlipsum}
\usepackage{verbatim}
\usepackage{xcolor}
\usepackage{CJK}

\newcolumntype{P}[1]{>{\centering\arraybackslash}p{#1}}

\begin{document}
\preprint{APS/123-QED}

\title{Zero-field magnetometry based on nitrogen-vacancy ensembles in diamond }
\author{Huijie Zheng}
\email{zheng@uni-mainz.de}
\affiliation{Johannes Gutenberg-Universit{\"a}t Mainz, 55128 Mainz, Germany}
\author{Jingyan Xu}
\affiliation{Chinese Academy of Sciences, Key Lab of Quantum Information, University of Science and Technology of China, Hefei 230026,
People's Republic of China}
\author{Geoffrey Z. Iwata}
\affiliation{Johannes Gutenberg-Universit{\"a}t Mainz, 55128 Mainz, Germany}
\author{Till Lenz}
\affiliation{Johannes Gutenberg-Universit{\"a}t Mainz, 55128 Mainz, Germany}
\author{Julia Michl}
\affiliation{Institute of Physics, University of Stuttgart and Institute for Quantum Science and Technology IQST, 70174 Stuttgart, Germany}
\author{Boris Yavkin}
\affiliation{Institute of Physics, University of Stuttgart and Institute for Quantum Science and Technology IQST, 70174 Stuttgart, Germany}
\author{Kazuo Nakamura} 
\affiliation{Application Technology Research Institute, Tokyo Gas Company, Ltd., Yokohama, 230-0045 Japan}
\author{Hitoshi Sumiya} 
\affiliation{Advanced Materials Laboratory, Sumitomo Electric Industries, Ltd., Itami, 664-0016 Japan}
\author{Takeshi Ohshima} 
\affiliation{Takasaki Advanced Radiation Research Institute, National Institutes for Quantum and Radiological Science and Technology, Takasaki, 370-1292, Japan }
\author{Junichi Isoya}
\affiliation{Faculty of Pure and Applied Sciences, University of Tsukuba, Tsukuba, 305-8573 Japan}
\author{J\"{o}rg Wrachtrup}
\affiliation{Institute of Physics, University of Stuttgart and Institute for Quantum Science and Technology IQST, 70174 Stuttgart, Germany}
\author{Arne Wickenbrock}
\affiliation{Johannes Gutenberg-Universit{\"a}t Mainz, 55128 Mainz, Germany}

\author{Dmitry Budker}
\affiliation{Johannes Gutenberg-Universit{\"a}t Mainz, 55128 Mainz, Germany}
\affiliation{Helmholtz Institut Mainz, 55099 Mainz, Germany}
\affiliation{Department of Physics, University of California, Berkeley, CA 94720-7300, USA}
\affiliation{Nuclear Science Division, Lawrence Berkeley National Laboratory, Berkeley, CA 94720, USA}

 \date{\today}

\begin{abstract}

Ensembles of nitrogen-vacancy (NV) centers in diamonds are widely utilized for magnetometry, magnetic-field imaging and magnetic-resonance detection. At zero ambient field, Zeeman sublevels in the NV centers lose first-order sensitivity to magnetic fields as they are mixed due to crystal strain or electric fields. In this work, we realize a zero-field (ZF) magnetometer using polarization-selective microwave excitation in a $^{13}$C-depleted crystal sample. We employ circularly polarized microwaves to address specific transitions in the optically detected magnetic resonance and perform magnetometry with a noise floor of 250\,pT/$\sqrt{\rm{Hz}}$. This technique opens the door to practical applications of NV sensors for ZF magnetic sensing, such as ZF nuclear magnetic resonance, and investigation of magnetic fields in biological systems.

\end{abstract}

\maketitle


\section{Introduction}

Negatively charged nitrogen-vacancy (NV) centers in diamond have garnered wide interest as magnetometers\,\cite{Wolf2015,Rondin2014,wickenbrock2016microwave,levelanticrossing,Blank2015,Cavity2}, with diverse applications ranging from electron spin resonance (ESR) and biophysics to materials science\,\cite{NVreview2006,nanozerofieldESR, Barry2016,balasubramanian2008nanoscale,NVImagingArray, Degen2008,London2013}.
However, typical operation of an NV magnetometer requires an applied bias magnetic field to nonambiguously resolve magnetically sensitive features in the level structure. Due to the Zeeman effect, the bias field lifts degeneracy among magnetic sublevels in the NV-center ground state, allowing microwave transitions between spin states to be addressed individually
\,\cite{SingleNVMagnetometry}. Such a bias field can be undesirable for applications where it could perturb the system to be measured, such as in magnetic susceptometry\,\cite{Eberbeck2009} and measurements in magnetically shielded environments, or, for example, can create challenging cross talk within sensor arrays\,\cite{Martinez2014}.

Elimination the need for a bias field would extend the dynamic range of NV magnetometers to zero field. Zero-field, NV-based magnetometry opens up new application avenues, and makes these versatile, solid-state sensors competitive with other magnetic field sensors such as superconducting quantum interference devices (SQUIDs) and alkali-vapor magnetometers\,\cite{PhysRevLett.120.033202,Quan17}, because, despite the lower sensitivity of NVs, they offer additional benefits due to their small size, high spatial resolution, capability of operation over large temperature and pressure ranges, and wide bandwidth\,\cite{Rondin2014}. The relative simplicity of NVs operated at zero field can readily complement existing sensors in applications such as zero- and ultralow-field nuclear magnetic resonance (ZULF-NMR)\,\cite{Jiangeaar6327,Ledbetter2008}, tracking field fluctuations in experimental searches for electric dipole moments\,\cite{Bison2018}, and magnetoencelography or magnetocardiography\,\cite{Ahonen1993,fenici2005}.

\begin{figure}[t]
  \centering
\includegraphics[width=\columnwidth]{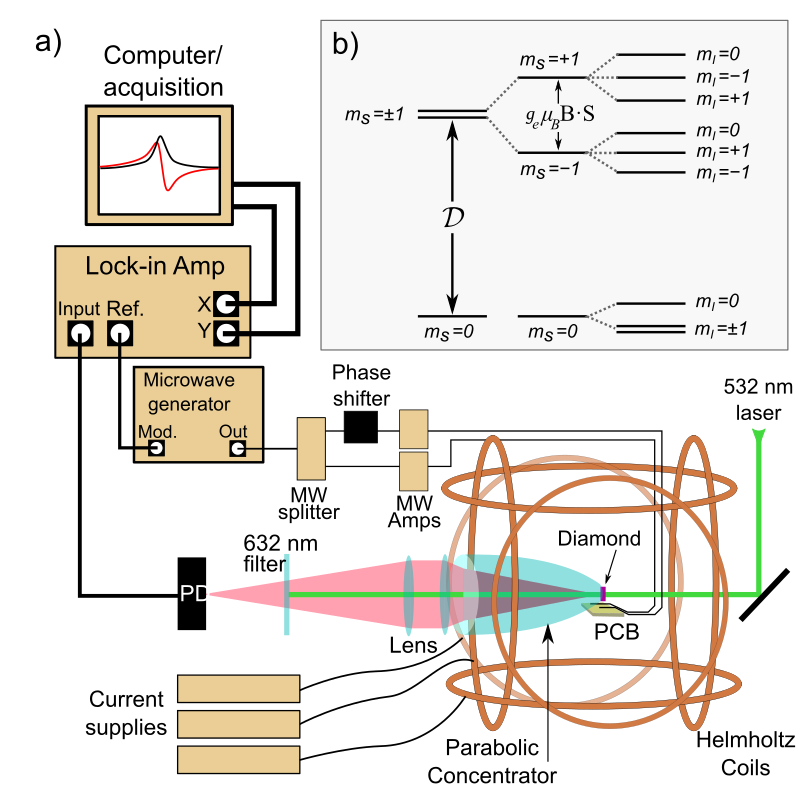}
\caption 
{ \label{fig:nvsetup} 
(a) The experimental setup for a zero-field NV magnetometer when modulating the MW frequency. PD: photodiode; MW: microwave; PCB: printed circuit board. (b) The ground-state-level diagram for NV in diamond. The label, $m_s$ refers to the electron spin projection, while $m_I$ refers to the nuclear spin projection.
}
\end{figure}

Magnetically sensitive microwave transitions within NV centers can be probed using the optically detected magnetic resonance (ODMR) technique, which relies on detecting changes in photoluminescence (PL) while applying microwave fields to optically pumped NVs\,\cite{ODMR1}. At zero field, these transitions overlap, and shift equally with opposite sign in response to magnetic fields.
Therefore, NV ensembles have been considered unusable as zero-field magnetometers \cite{Rondin2014}, except in certain cases, for detecting ac fields in the presence of applied microwaves\,\cite{ACmagtometerJapan2018}.

We overcome these complications at zero field by selectively driving resolved hyperfine transitions in NV centers in a $^{13}$C-depleted diamond with frequency-modulated, circularly polarized microwaves. This results in ODMR fluorescence with a linear response to small magnetic fields. We present such a zero-field NV magnetometer with a demonstrated noise floor of 250\,pT/$\sqrt{\mathrm{Hz}}$. 

\begin{figure*}[t]
  \centering
\includegraphics[width=1.8\columnwidth]{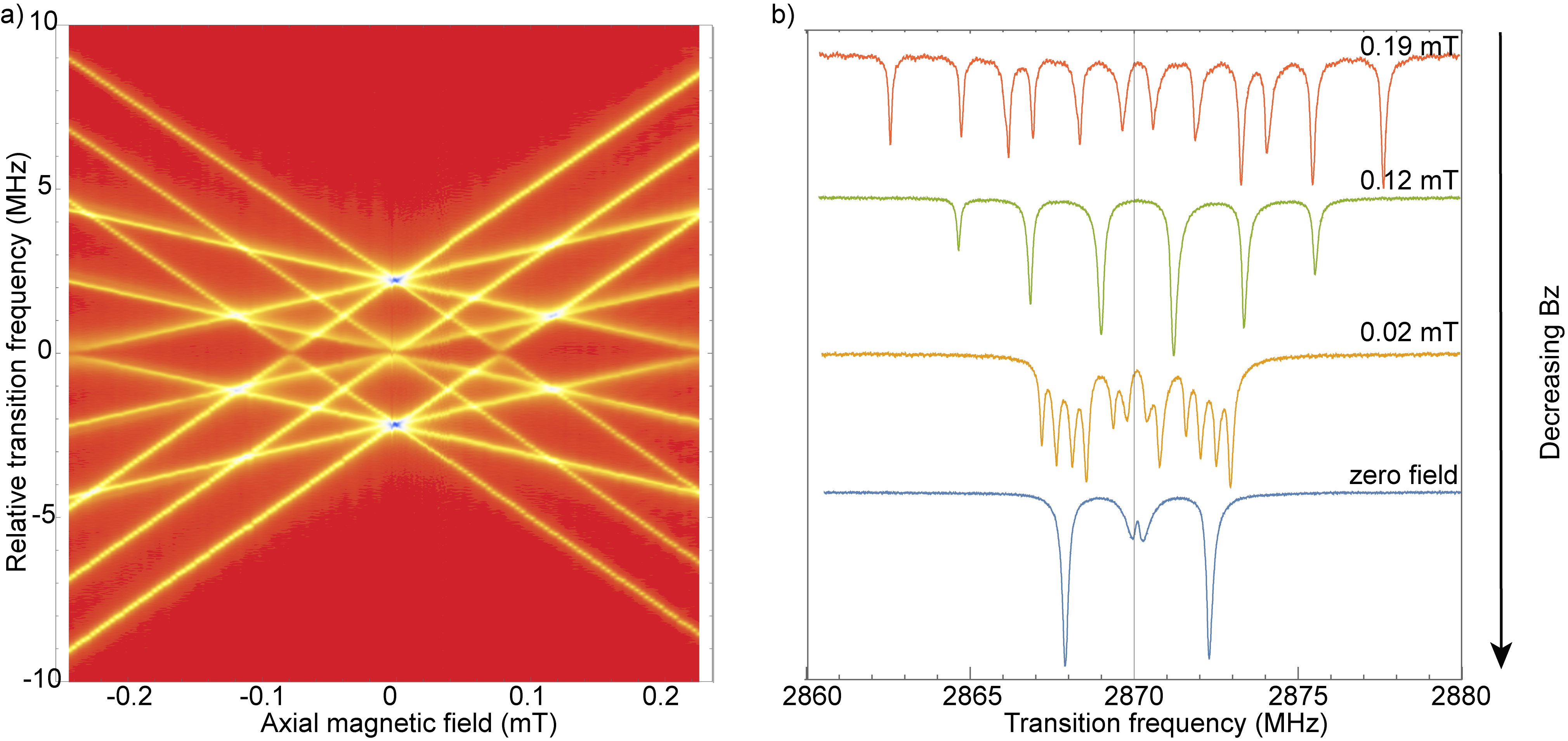}
\caption 
{\label{Fig:ODMR} 
(a) ODMR spectra with linearly polarized MWs as a function of the axial magnetic field, with transitions originating from all crystal-axis orientations. Those transitions corresponding to NVs oriented along the direction of the applied field are labeled ($|m_s,m_I\rangle$) and overlaid with the simulation according to the Hamiltonian of Eq.\,\ref{eq:A1}. The unlabeled transitions are due to NV centers that are not oriented along the $\langle 111 \rangle$ direction. These features from other orientations overlap because they share a common relative angle with the applied field. (b) Continuous-wave ODMR spectra for selected values of $B_z$. At zero field only the central transitions are split. The lower-energy peak at $\approx 2868$\,MHz corresponds to the transitions $|m_s=0,m_I=+1\rangle \rightarrow |m_s=+1,m_I=+1\rangle$ and $|m_s=0,m_I=-1\rangle \rightarrow |m_s=-1,m_I=-1\rangle$. The higher-energy peak at $\approx 2872$\,MHz corresponds to the transitions $|m_s=0,m_I=-1\rangle \rightarrow |m_s=+1,m_I=-1\rangle$ and $|m_s=0,m_I=+1\rangle \rightarrow |m_s=-1,m_I=+1\rangle$.}
\end{figure*}

\section{Experiment}

A schematic of the experimental apparatus is shown in Fig.\,\ref{fig:nvsetup}\,(a). Intensity stabilized ($<$0.5\% power fluctuations at 110 mW) green laser light at 532 nm is used to optically pump the NV centers in diamond into a single spin projection ($m_s=0$) in the ground state. Microwaves (MWs) drive transitions from this ground state into the magnetically sensitive spin states ($m_s=\pm1$), reducing the fluorescence resulting from the pumping cycle \cite{Jensen2017}. The diamond is glued to a parabolic light concentrator to collect fluorescence, which is focused through a filter and onto a photodetector (PD), which registers $\approx 1.2$ mW. The parabolic-concentrator arrangement has been demonstrated to have over 60\% collection efficiency in Ref.\,\cite{Wolf2015}. The signal from the PD is fed into a lock-in amplifier (LIA), which is referenced to the frequency modulation of the MWs. Three pairs of Helmholtz coils are wound onto a three-dimensionally (3D) printed mount around the diamond sample. To zero the ambient magnetic field, the coils are driven by three independent stable current supplies.


The sensor is a 99.97\% $^{12}$C, (111)-cut diamond single crystal, with dimensions $0.71\,\mathrm{mm}\times 0.69\,\mathrm{mm}$ and a thickness of $0.43\,\mathrm{mm}$. It is laser-cut from a $^{12}$C-enriched diamond single crystal grown by the temperature-gradient method at high pressure (6.1\,GPa) and high temperature ($1430^{\mathrm{o}}$C). A metal solvent containing a nitrogen getter and carbon powder prepared by pyrolysis of 99.97\% $^{12}$C-enriched methane as a carbon source were used\,\cite{Nakamura2007}. It was irradiated with 2 MeV electrons from a Cockcroft-Walton accelerator to a total fluence of $1.8\times10^{18} \mathrm{cm}^{-2}$ at room temperature, and annealed at 800$^{\mathrm{o}}$C for 5 hours. This source diamond was reported to have 3-ppm initial nitrogen and 0.9-ppm NV$^-$ after conversion, measured by electron-paramagnetic-resonance techniques \cite{Wolf2015}.


To study the ODMR signal around zero field, fluorescence spectra are taken by sweeping the center frequency of linearly polarized MWs for a range of magnetic field values around zero. Figure \ref{Fig:ODMR}\,(a) shows the resulting data, which show hyperfine resolved transitions originating from all crystal orientations. The narrow linewidths in this diamond allow for a clear distinction between transitions occurring along different crystal axes. These transitions are shown in detail at specific field values in Fig.\,\ref{Fig:ODMR}\,(b), including at zero field, where 12 transitions overlap and merge into four distinct features.


The NV spin Hamiltonian that describes the energy spectrum, and includes interaction with an applied magnetic field $\mathbf{B}$ and hyperfine interaction with the intrinsic $^{14}$N is written as follows
\begin{equation}
\begin{split}
\mathcal{H}= \mathcal{D}\mathbf{S}_{\text{z}}^{\textrm{2}}&+\it{E}(\mathbf{S}_{x}^\textrm{2}-\mathbf{S}_{y}^\textrm{2})+\it{g}_{\it{e}}\mu_{B}\mathbf{B}\cdot\mathbf{S}\\ & +\mathbf{S}\cdot\mathbf{A}\cdot\mathbf{I}-\it{g}_{\it{n}}\mu_{N}\mathbf{B}\cdot\mathbf{I}+\it{Q}\mathbf{I}_{\text{z}}^{\textrm{2}},\label{eq:A1}
\end{split}
\end{equation}
where $\mathcal{D}=2870$\,MHz is the zero-field-splitting parameter, $\it{E}$ ($\approx$ 0.15\,MHz for the employed sample) is the off-diagonal strain- or electric field-splitting tensor\,\cite{PhysRevLett.121.246402}, $\mathbf{S}$ and $\mathbf{I}$ are the electron- and nuclear-spin operators, $\it{g}_{\it{e}}$ and $\it{g}_{\it{N}}$ are the electron and nuclear-spin \emph{g}-factors, $\mu_{B}$ and $\mu_{N}$ are the Bohr and nuclear magnetons, and $\it{Q}=-5$\,MHz is the nuclear-quadrupole-splitting parameter. The electron spin $\mathbf{S}$ and nuclear spin $\mathbf{I}$ are coupled via the diagonal hyperfine tensor
\begin{equation}
\mathbf{A}=
\begin{pmatrix}
    \it{A}_{\bot}  & 0 & 0 \\
    0  &\it{A}_{\bot} &0 \\
    0  & 0 &\it{A}_{\parallel}\\
\end{pmatrix},\label{eq:eqA2}
\end{equation}
where $\it{A}_{\bot}$=-2.7\,MHz and $\it{A}_{\parallel}$ =-2.16\,MHz \cite{PhysRevB.79.075203}. To determine the parameter $\it{E}$ for the sample, the experimentally measured spectra are fitted to the Hamiltonian with $\it{E}$ as a variable and other parameters fixed to values reported in the literature.

At low fields, the Hamiltonian is dominated by the 2870-MHz zero-field splitting, $\mathcal{D}$, between the $|m_s=0\rangle$ and the degenerate $|m_s=\pm1\rangle$ states. The hyperfine interaction with the nuclear spin of the NVs' intrinsic $^{14}$N results in three hyperfine projections for each electron spin state, and so the MW transitions between the $m_s$ states are split threefold [Fig.\,\ref{fig:nvsetup} (b)]. These groupings of $|m_s=\pm1\rangle$ states separate in energy with increasing magnetic field due to the Zeeman effect. At fields where the Zeeman shift results in degenerate hyperfine levels of the $|m_s=\pm 1\rangle$ states, anticrossings between hyperfine states with the same nuclear spin projection occur due to the tensor $\it{E}$\,\cite{PhysRevLett.121.246402}. 

These anticrossings between hyperfine states are apparent in the ODMR Zeeman spectra of Fig.\,\ref{Fig:ODMR}\,(a). The anticrossings for the transition energies at $\approx \pm 0.08$\,mT correspond to interaction between the states $|m_s=-1,m_I=\pm1\rangle$ and $|m_s=+1,m_I=\pm1\rangle$. The bottom spectrum in Fig.\,\ref{Fig:ODMR}\,(b) shows strain- and electric field- splitting of the transitions $|m_s=0,m_I=0\rangle \rightarrow |m_s=\pm1,m_I=0\rangle$ due to the interaction $E$ between the upper states, while the transitions $|m_s=0,m_I=+1\rangle \rightarrow |m_s=\pm1,m_I=+1\rangle$, and $|m_s=0,m_I=-1\rangle \rightarrow |m_s=\mp1,m_I=-1\rangle$ merge in the ODMR spectrum. In many diamond samples these features are indiscernible, because the transverse zero-field splitting $E$ is larger than the intrinsic hyperfine splitting of the NV center. The well-resolved hyperfine structure in this diamond sample allows us to selectively address only the overlapping transitions that occur at $\pm 2$\,MHz from the central feature.

Note that here we use the transitions that overlap at either $\sim$2868\,MHz or $\sim$2872\,MHz for zero-field magnetometry, since the splitting, $E$, from the strain and/or electric field is suppressed as a result of being at least an order of magnitude smaller than the energy separation between the corresponding branches, $|m_s=\pm1,m_I=+1\rangle$ or $|m_s=\pm1,m_I=-1\rangle$, which are split by the hyperfine-coupling term. These features are also used to calibrate the zero-field value of currents in our coils, by maximizing their fluorescence contrast, which occurs when the magnetic sublevels become degenerate as a result of the external magnetic field being zeroed. The estimated residual field is less than 3\,$\mu$T in all directions which corresponds to the linewidth of these transitions.

\begin{figure}
  \centering
\includegraphics[width=\columnwidth]{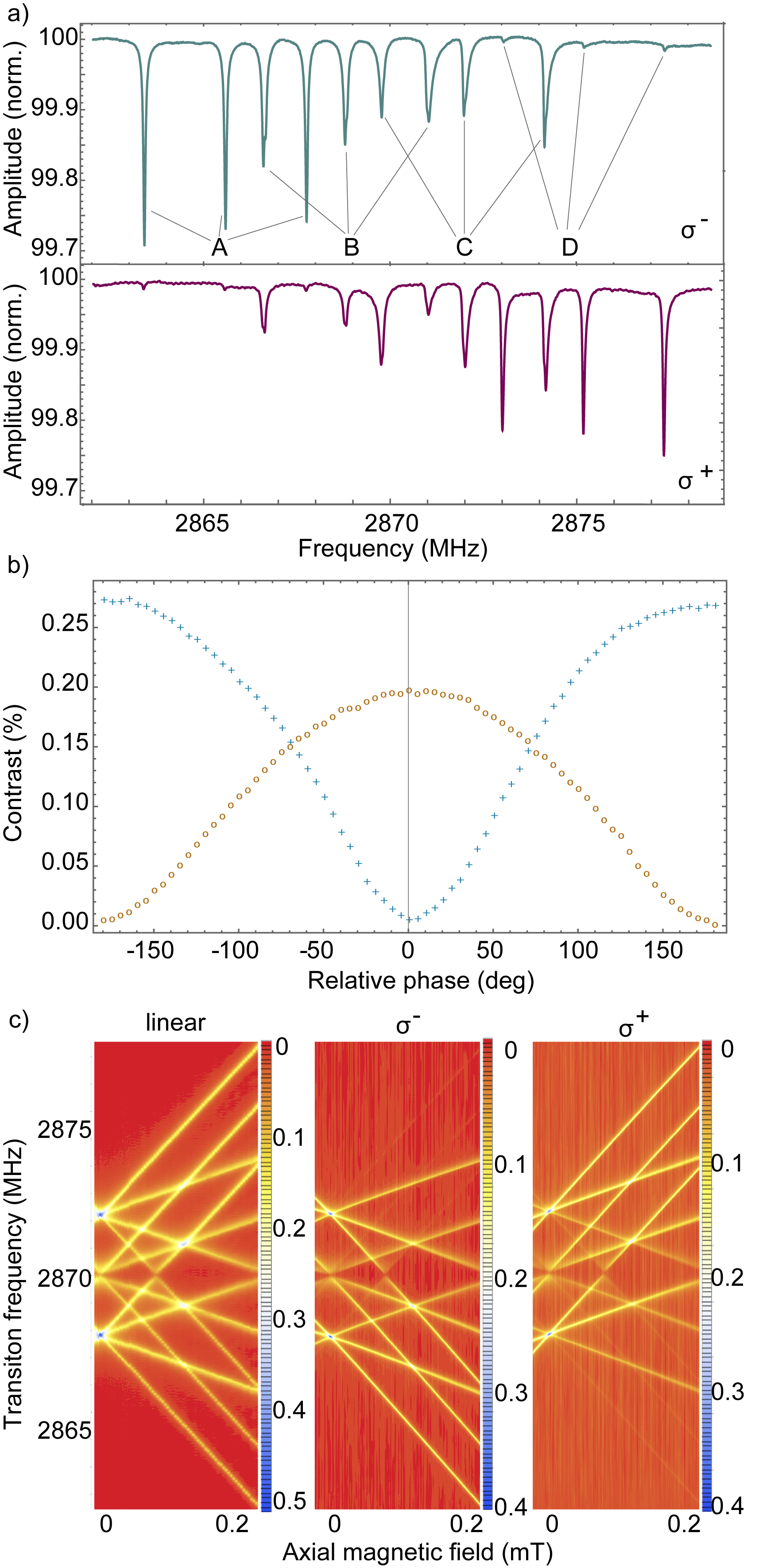}
\caption 
{ \label{Fig:fieldmod} 
(a) ODMR traces at fixed field with circular MWs (the polarization of the applied MW is indicated at the right bottom corner of each subfigure: top $\sigma^-$ and bottom $\sigma^+$). Here the peaks $A$ and $D$ ($B$ and $C$) correspond to the transitions from $|m_s=0\rangle$ to $|m_s=-1\rangle$ and $|m_s=0\rangle$ to $|m_s=1\rangle$ of on-axis (off-axis) NVs, respectively. (b) Fitted amplitudes of $A$ and $D$ in (a), as a function of the relative phase between two applied microwave fields. Blue crosses (amber circles) indicate the amplitudes of $A$ ($D$). The error bars on the data (±1 standard error of the mean) are smaller than the symbol size. (c) ODMR spectra under linear, $\sigma^-$, $\sigma^+$ MW as a function of magnetic field. The color scale indicates the peak depth, in percent, relative to the off-resonant case.}
\end{figure}

We use circularly polarized MWs to drive a single electron spin transition out of the feature composed of overlapping resonances. This single transition has a linear dependence on the magnetic field. The circularly polarized MWs are created using a printed circuit board (PCB) that follows the design of Ref.\,\cite{Mrozek2015}. The board consists of two, 200-$\mu$m wires separated by a distance $d=4.5$\,mm and placed on a plane that is $d/2$ away from the diamond sample. Each wire carries a MW signal split from the same source, with one passed through a variable phase shifter. This arrangement results in orthogonal oscillating magnetic fields at the diamond sample, verified with the high contrast between ODMR traces at fixed field with either left or right circularly polarized MWs [Fig.\,\ref{Fig:fieldmod}\,(a)]. The MW field can be continuously varied between linearly and circularly polarized [Fig.\,\ref{Fig:fieldmod}\,(b)].

The efficacy of our circular MW polarizing scheme is shown in Fig.\,\ref{Fig:fieldmod}. In particular, Fig.\,\ref{Fig:fieldmod}\,(b,c) demonstrate a relative suppression of $\sigma^+$ and $\sigma^-$ transitions to below 1\% of the maximum contrast, respectively. Previously overlapping transitions are thus isolated, removing the symmetric dependence as a function of the field to be measured.


\begin{figure*}
  \centering
\includegraphics[width=1.8\columnwidth]{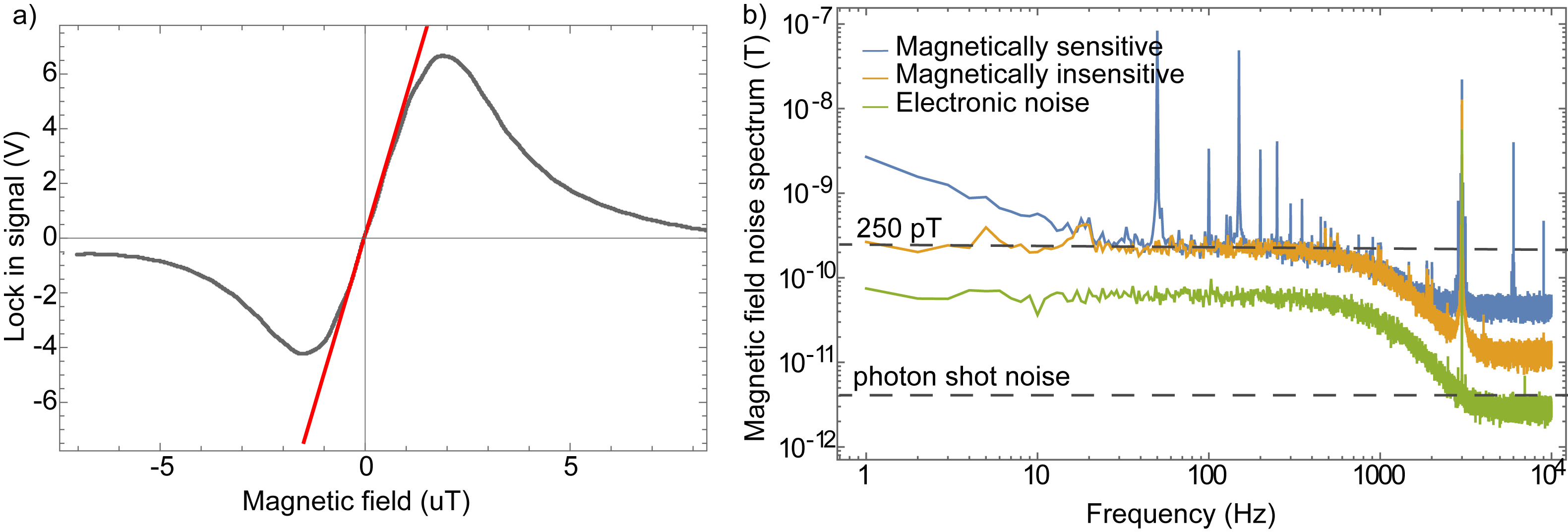}
\caption 
{ \label{Fig:freqmod} 
(a) A detail of the LIA output $X$ around zero field (black line) with a dispersive curve fitting and a linear fit (red) to the data while modulating the frequency of the applied circularly polarized microwave at 3\,kHz with a depth of 45\,kHz. (b) The magnetic-field noise spectrum. The blue line indicates magnetically sensitive noise, the amber line indicates magnetically insensitive at a MW frequency of $\approx$2900\,MHz (average noise between 1 and 1000\,Hz is 250\,pT/$\sqrt{\rm{Hz}}$), and the green line indicates electronic noise (average noise between 1 and 1000\,Hz is 70\,pT/$\sqrt{\rm{Hz}}$). The photon shot-noise limit of the magnetometer is indicated at 4\,pT/$\sqrt{\rm{Hz}}$. The decrease in signal for frequencies above 1\,kHz is due to the filtering of the LIA. }
\end{figure*}


\section{Magnetometry Method}

In typical ODMR magnetometry, linearly polarized MW fields drive transitions between the $|m_s=0\rangle$ and $|m_s=\pm 1\rangle$ states, decreasing the detected fluorescence with a resonant response with respect to the MW frequency. At fields where the transitions to $|m_s=\pm 1\rangle$ states are well resolved, modulation is applied to the MW frequency, and the resulting PL signal is detected on a PD and demodulated on a LIA at the modulation frequency. The first-order harmonic output exhibits a linear response of the PL to the magnetic field. However, at zero field, the $|m_s=\pm 1\rangle$ states are degenerate, and transitions to these states, including those from all crystal orientations, overlap, causing the LIA output to no longer exhibit a linear dependence with the magnetic field, as described below. 

In our experimental setup, we measure a field that is applied along the $\langle 111 \rangle$ direction. While the angle between the $\langle 111 \rangle$ oriented NVs and the applied field is 0$^\mathrm{o}$, the other three NV orientations are all oriented with a 70$^\mathrm{o}$ angle to the applied field. Therefore, the applied field is projected onto the three other crystal orientations equally, but a single-axis description is sufficient to understand how magnetic field sensitivity arises in the $\langle 111 \rangle$ oriented NVs.

For a background PL of $P_0$, the detected signal from applying MWs at frequency $\omega$ to a given transition can be modeled as follows 
\begin{equation}
   P=P_0 - \frac{A\left(\frac{\Gamma}{2}\right)^2}{(\omega-\omega_{0})^2+\left(\frac{\Gamma}{2}\right)^2},
   \label{eq:pl}
\end{equation}
where $P$ is the PL signal, and $\omega_{0}$, $\Gamma$, and $A$ are the center frequency, the linewidth and the amplitude of this Lorentzian profile. 

At zero field, the PL signal is convolved with features due to strain and electric field, however, we can still approximate the signal with the Lorentzian form in Eq.\,\ref{eq:pl}. Focusing on the lowest-frequency peak of those shown in the zero-field trace of Fig.\,\ref{Fig:ODMR} (b), two transitions, $|m_s=0,m_I=+1\rangle \rightarrow |m_s=+1,m_I=+1\rangle$ with amplitude $A_+$ and central frequency $\omega_{0+}$, and $|m_s=0,m_I=-1\rangle \rightarrow |m_s=-1,m_I=-1\rangle$ with amplitude $A_-$ and central frequency $\omega_{0-}$, overlap such that $\omega_{0+}=\omega_{0-}=\omega_{0}$. If the field $B_z$ along a single diamond axis is increased, the central frequencies change by an amount,
\begin{equation}
\omega_{0\pm}=\omega_0\pm \it{g}_{\it{e}}\mu_{B} B_z.
\label{eq:del-omega}
\end{equation}
The effect of a small modulation of the MW frequency is described by the first-order expansion of $P$ for each transition around $\omega_0$ with respect to $\omega$. The result is the linear dependence on the magnetic field, and is a sum of the contributions from each transition. Using the relation in Eq.\,\ref{eq:del-omega}, we find that for small values of $B_z$,
\begin{equation}
\frac{\Delta P}{\Delta \omega} =(A_+ - A_-)K B_z+c,
\label{eq:freq}
\end{equation}
where we group various terms into the parameters $K=8 g_{e}\mu_{B}/\Gamma^2$ and $c=-8 (\omega-\omega_0)(A_++A_-)/\Gamma^2$. When linearly polarized microwaves are applied, the transition probabilities for each $m_s$ state, and therefore the values of $A_\pm$, are equal. As a result, the linear change in PL is zero for small changes in magnetic field. However, circularly polarized microwaves can be applied instead, resulting in different transition probabilities so that the first term in Eq.\,\ref{eq:freq} does not cancel out, resulting in magnetically sensitive changes in the PL. Therefore, to perform high-sensitivity magnetometry, we apply circularly polarized MWs and modulate the central frequency. 
Note here that the shape of the magnetometry signal is sensitive to the imperfection of circular MW polarization, which has varying effects for other crystal orientations, and to small detunings of the central MW frequency, which may arise from drifts in the diamond temperature. These effects can explain the asymmetry in Fig.\,\ref{Fig:freqmod}\,(a).

The contributions of the NVs oriented along other crystallographic axes also overlap in the lineshape but they have a weaker dependence on the magnetic field (applied at an angle to these axes) and are suppressed when circularly polarized MWs are applied. As a result, the linear dependence described above is the dominant behavior for small fields.

\section{Alternative Magnetometry Method}

As an aside, we mention that it is also possible to apply an oscillating magnetic field along the $\langle 111 \rangle$ direction in order to perform magnetometry with linearly polarized MWs. This oscillating field modulates the central frequencies of the two transitions as follows
\begin{equation}
\omega_{0\pm}=\omega_0 \pm \left[\it{g}_{\it{e}}\mu_{B} B_z+ \omega_{mod} \right],
\label{eq:del-field}
\end{equation}
where $\omega_{mod}=g_{e}\mu_{B}\eta\,\mathrm{sin}(\nu t)$ and $\eta$ and $\nu$ are the amplitude and frequency of the oscillating field, respectively. The resulting PL is the sum of contributions from the relevant transitions, each described by the first-order expansion of Eq.\,\ref{eq:pl} in $\omega_{mod}$ around $\omega_{mod}=0$. The linear magnetic dependence for small values of $B_z$, and $\eta\ll\Gamma/g_{e}\mu_{B}$ is,  
\begin{equation}
\frac{\Delta P}{\Delta \omega_{mod}} = (A_+ + A_-)\, K B_z +c^\prime,
\label{eq:field}   
\end{equation}
where $K$ is as defined in Eq. \ref{eq:freq} and $c^\prime=8 (\omega-\omega_0)(A_--A_+)/\Gamma^2$. The signal is detected on a LIA referenced to the field modulation frequency, $\nu$. Since $A_\pm$ do not cancel out in Eq.\,\ref{eq:field}, there is linear magnetic sensitivity in the LIA output signal, which, for $A_-=A_+$, is insensitive to the detuning of the central MW frequency within the linear regime. The magnetic dependence, $\Delta P/\Delta B_z$, reaches a maximum when $\eta=\Gamma/2$.

This method can be simpler to implement in some applications for dc measurements of small fields, since compensation coils, for instance, can be used to apply the modulation, and there is no need for circularly polarized microwaves. In certain dc and low-frequency applications, such as for biomagnetic signals, this field modulation can be averaged out. Furthermore, the employment of an oscillating bias field relaxes the constraints on bias stability, a concern for precision sensors. 


\section{Magnetic field sensitivity}

To demonstrate the magnetic sensitivity of the zero-field magnetometer, we scan the magnetic field through zero while modulating the frequency of the $\sigma^+$-polarized MWs, which are centered at $\sim$ 2872 MHz. The derivative fluorescence signal as detected in a properly phased LIA output depends linearly on the field between $\approx \pm 1\,\mu$T, which determines the dynamic range of the magnetometer (which can be extended by applying magnetic bias). The calibration signal is shown in Fig.\,\ref{Fig:freqmod}\,(a) with a modulation frequency of 3\,kHz, and a modulation depth of 45\,kHz. The data near zero field are fitted to a straight line, and the slope of this line is used to translate the LIA output signal to the magnetic field. The noise in the LIA output signal voltage therefore conveys the sensitivity of the magnetometer. For noise measurements, the LIA output is recorded for 1\,s while the background magnetic field is set to zero. The data are passed through a fast Fourier transform and displayed in Fig.\,\ref{Fig:freqmod}\,(b), from which we can establish our noise floor and thus the sensitivity, for a given bandwidth.

For noise frequencies between dc-30\,Hz we observe a $1/f$-behavior of the magnetic noise that we attribute to ambient noise, primarily arising from the compensation-coil current stability. While the noise floor at 250\,pT$/\sqrt{\rm{Hz}}$ can be attributed to laser-intensity noise, a photon shot-noise limit of 4\,pT/$\sqrt{\rm{Hz}}$ is achievable---a value calculated from the number of incident photons at the photodetector and the contrast of the magnetically sensitive feature in the MW spectral response. The magnetically insensitive noise spectrum is obtained by operating the setup at an off-resonant microwave frequency of 2900\,MHz, where there are no magnetically sensitive features in the NV spectrum. Since noise peaks at 50\,Hz and harmonics are absent in this magnetically in-sensitive spectrum, we attribute them to magnetic noise in the laboratory.
The electronic noise floor ($\approx70$\,pT/$\sqrt{\rm{Hz}}$) was measured by turning off the green excitation light and acquiring the output of the LIA.

\section{Conclusion}

We demonstrate a NV-based zero-field magnetometer with a 250\,pT$/\sqrt{\rm{Hz}}$ noise floor. The device employs a diamond sample with a well-resolved ground-state hyperfine structure of the NV center, and uses circularly polarized microwaves to selectively excite magnetically sensitive transitions that, in the absence of such selectivity, yield ODMR signals that are first-order insensitive to near-zero magnetic fields. This device can be useful in applications where a bias field is undesirable and extends over the dynamic range of NV magnetometry to cover existing zero-field technologies such as SQUIDs and alkali-vapor magnetometers. ZF magnetometry with single NVs will be presented in a later publication. Improvements in the present technique will result in sensitivities that are useful for ZULF-NMR and, with further miniaturization, these zero-field diamond sensors can find use in biomagnetic applications such as magnetoencelography and magnetocardiography.

\section{Acknowledgements}

We gratefully acknowledge W. Gawlik for providing an early version of the circuit board for producing circularly polarized microwaves. We thank G. Balasubramanian, Philipp Neumann, A. Jarmola, G. Chatzidrosos and J. W. Blanchard for informative discussions and fruitful advice. This work was supported by the EU FET-OPEN Flagship Project ASTERIQS (Action 820394) and the German Federal Ministry of Education and Research (BMBF) within the Quantumtechnologien program (FKZ 13N14439) and the Japan Society of the Promotion of Science (JSPS) KAKENHI (Grant No.17H02751).

\bibliographystyle{apsrev4-2-2}
\bibliography{literature,literatureg}

\end{document}